\documentclass[aps,prb,twocolumn,superscriptaddress]{revtex4}

\usepackage{graphicx}
\usepackage{subfigure}

\begin{document}


\title{Non-local transport in normal-metal/superconductor hybrid structures: the role of interference and interaction}


\author{J. Brauer}
\affiliation{Karlsruher Institut f\"ur Technologie, Institut f\"ur Nanotechnologie, Karlsruhe, Germany}
\author{F. H\"ubler}
\affiliation{Karlsruher Institut f\"ur Technologie, Institut f\"ur Nanotechnologie, Karlsruhe, Germany}
\affiliation{Karlsruher Institut f\"ur Technologie, Center for Functional Nanostructures, Karlsruhe, Germany}
\affiliation{Karlsruher Institut f\"ur Technologie, Institut f\"ur Festk\"orperphysik, Karlsruhe, Germany}
\author{M. Smetanin}
\affiliation{Karlsruher Institut f\"ur Technologie, Institut f\"ur Nanotechnologie, Karlsruhe, Germany}
\author{D. Beckmann}
\email[e-mail address: ]{detlef.beckmann@kit.edu}
\affiliation{Karlsruher Institut f\"ur Technologie, Institut f\"ur Nanotechnologie, Karlsruhe, Germany}
\affiliation{Karlsruher Institut f\"ur Technologie, Center for Functional Nanostructures, Karlsruhe, Germany}
\author{H. v. L\"ohneysen}
\affiliation{Karlsruher Institut f\"ur Technologie, Center for Functional Nanostructures, Karlsruhe, Germany}
\affiliation{Karlsruher Institut f\"ur Technologie, Institut f\"ur Festk\"orperphysik, Karlsruhe, Germany}
\affiliation{Karlsruher Institut f\"ur Technologie, Physikalisches Institut, Karlsruhe, Germany}


\date{\today}

\begin{abstract}
We have measured local and non-local conductance of mesoscopic normal-metal/superconductor hybrid structures fabricated by e-beam lithography and shadow evaporation. The sample geometry consists of a superconducting aluminum bar with two normal-metal wires forming tunnel contacts to the aluminum at distances of the order of the superconducting coherence length. We observe subgap anomalies in both local and non-local conductance that quickly decay with magnetic field and temperature. For the non-local conductance both positive and negative signs are found as a function of bias conditions, indicating at a competition of crossed Andreev reflection and elastic cotunneling. Our data suggest that the signals are caused by a phase-coherent enhancement of transport rather than dynamical Coulomb blockade.
\end{abstract}


\maketitle


\section{Introduction}
Andreev reflection\cite{andreev1964} (AR) is the process responsible for the transfer of electrons from a normal metal into a superconductor at energies below the superconducting gap energy. In multi-terminal structures with two (or more) normal metals (N) connected to a single superconductor (S), non-local or crossed Andreev reflection (CAR) may occur, where an electron entering the superconductor from one normal-metal contact A forms a Cooper pair by emitting a hole into a second contact B.\cite{byers1995,deutscher2000} A competing process is elastic cotunneling (EC),\cite{falci2001} where an electron is transmitted to contact B without the formation of a Cooper pair. CAR has attracted theoretical and experimental\cite{beckmann2004,beckmann2007,russo2005,cadden2006,cadden2007,cadden2009,asulin2006,kleine2009,hofstetter2009} attention mainly because it is predicted to create spatially separated, entangled electrons in a solid-state environment (see Ref.~\onlinecite{burkard2007} for a brief review). While CAR can be readily probed by spin selection\cite{beckmann2004,beckmann2007} using ferromagnetic electrodes, this approach is unsuitable for entangler devices, since projecting the spin will destroy entanglement. Therefore, understanding the competition between CAR and EC in multiterminal NSN structures is a prerequisite for the successful design of superconducting solid-state entanglers. In a recent experiment a bias-dependent crossover from EC to CAR has been observed in a diffusive NSN structure with low-transparency tunnel contacts.\cite{russo2005} While both quantum mechanical interference \cite{russo2005,morten2006,melin2006,duhot2006,brinkman2006} and Coulomb interaction \cite{yeyati2007,kleine2009} have been proposed to explain the result, its origin is not yet clear. Here, we present an experimental investigation of both local and non-local transport in lateral NSN hybrid structures. We observe a subgap anomaly in the non-local conductance similar to the observations in Ref.~\onlinecite{russo2005}. Comparison with local-transport data reveals that the non-local signal is controlled by a phase-coherent enhancement of local transport. Dynamical Coulomb blockade is shown to be present in the samples, but can be clearly ruled out as the cause of the non-local conductance signal by its different dependence on temperature and magnetic field.

\section{Experimental}
\begin{figure}
\includegraphics[width=\columnwidth]{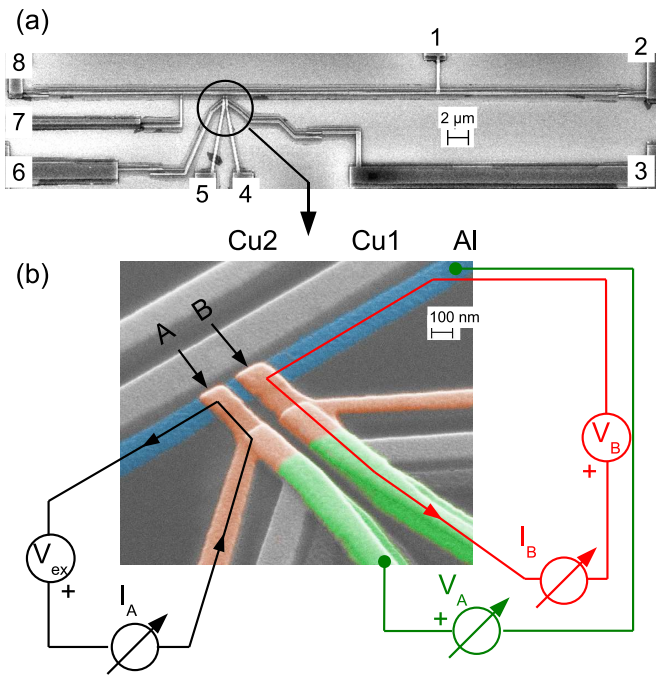}
\caption{\label{fig_sem}(Color online) (a) SEM image of the sample layout. (b) Close-up of the contact region, colorized for clarity. Two copper (orange) wires (A and B) are connected by overlap tunnel junctions to a weakly oxidized aluminum bar (blue). The copper wires fork near the tunnel contact to allow four-probe characterization. As a by-product of shadow evaporation, additional unconnected parts of the sample (grey), and a Cu/Al/oxide/Cu trilayer (green), are formed. The contact configuration and measurement scheme is also indicated. The three metal layers (Cu2, Cu1, Al) as created by shadow evaporation are indicated above the image.}
\end{figure}

Figure \ref{fig_sem}(a) shows an SEM image of the layout of our samples. They mainly consist of a narrow superconducting aluminum wire of about $60~\mathrm{\mu m}$ length (contacts 1, 2 and 8), and perpendicular to it, two copper wires (contacts 3/4 and 5/6) forming tunnel junctions A and B to the aluminum. A third tunnel junction (contact 7) was used for some control experiments. The central junction area is shown on an enlarged scale in panel (b), together with a typical measurement configuration. Three metal layers were evaporated onto an oxidized silicon substrate under different angles through a shadow mask fabricated by standard e-beam lithography technique. First, an auxiliary layer of 30~nm thick copper (Cu1), was evaporated to form Ohmic contacts to the two subsequent layers at  interconnects to the lines 2,3,6,7 and 8. In the second evaporation step under a different angle, the superconductor strip, an aluminum bar of 30~nm thickness, was created. The aluminum film was then oxidized {\em in situ} by applying $\approx 1~$Pa of pure oxygen for $\approx 5~$min to form a tunnel barrier. Then the third layer (Cu2), made of copper with thickness $t_\mathrm{Cu2}$, was evaporated, forming the two tunnel contacts A and B to the aluminum. Sample parameters such as contact areas, resistivities, etc., varied slightly between fabrication batches. Consistent results were obtained from six different samples, with parameters given in Table \ref{tab_params}. We show here mainly data from samples I and II.

\begin{table}
\caption{\label{tab_params}Characteristic parameters of the six samples.
Film thickness $t_\mathrm{Cu2}$, contact distance $d$, coherence length $\xi$, aluminum resistivity $\rho_\mathrm{Al}$ at $T=4.2~\mathrm{K}$, 
normal-state tunnel junction conductances $G_\mathrm{A}$ and $G_\mathrm{B}$, and contact transparencies $\mathcal{T}_\mathrm{A}$ and $\mathcal{T}_\mathrm{B}$.}
\begin{ruledtabular}
\begin{tabular}{lcccccccc}
sample & $t_\mathrm{Cu2}$ & $d$  & $\xi$ & $\rho_\mathrm{Al}$         & $G_\mathrm{A}$ & $G_\mathrm{B}$        & $\mathcal{T}_\mathrm{A}$     & $\mathcal{T}_\mathrm{B}$ \\ 
       & (nm)             & (nm) & (nm)  & ($\mathrm{\mu \Omega cm}$) & \multicolumn{2}{c}{$(\mathrm{\mu S)}$} & \multicolumn{2}{c}{$(\times 10^{-5})$} \\ \hline
I      & 20               & 75   & 120   & 4.1                       & 1105 & 1530 & 4.9               & 5.6   \\
II     & 20               & 100  & 120   & 4.2                       & 660  & 760  & 5.3               & 5.2   \\
III    & 20               & 115  & 170   & 2.1                       & 520  & 380  & 4.6               & 3.1   \\
IV     & 30               & 300  & 140   & 3.1                       & 1020 & 1150 & 3.3               & 3.3   \\
V      & 30               & 105  & 110   & 5.1                       & 1480 & 1520 & 7.7               & 7.4   \\
VI     & 30               & 105  & 110   & 5.3                       & 1390 & 1110 & 7.3               & 5.6   \\
\end{tabular}
\end{ruledtabular}
\end{table}

The samples were mounted into a shielded box thermally anchored to the mixing chamber of a dilution refrigerator. The measurement lines were fed through a series of filters to eliminate RF and microwave radiation from the shielded box.
A voltage $V_\mathrm{ex}$ consisting of a variable DC bias and a low-frequency AC excitation was applied to contact A, and the resulting currents $I_\mathrm{A}$ and $I_\mathrm{B}$ through both contacts were measured using independent current amplifiers. The actual voltage $V_\mathrm{A}$ across  the contact was measured via additional leads in a four-probe configuration. Interchanging current and voltage probes did not change the observed signals. Measurements were also performed with the roles of injector and detector interchanged between contact A and B, with consistent results. In some cases, an additional DC bias $V_\mathrm{B}$ was applied to contact B. Voltage and current polarities are indicated in Fig.~\ref{fig_sem}(b) by plus signs and arrows, respectively. For our choice of current polarities EC leads to a positive non-local conductance, while CAR leads to a negative signal. The AC components of $V_\mathrm{A}$, $I_\mathrm{A}$ and $I_\mathrm{B}$ were measured simultaneously with lock-in technique, and from the in-phase signals the local and non-local differential conductances $g_\mathrm{AA}=dI_\mathrm{A}/dV_\mathrm{A}$ and $g_\mathrm{AB}=dI_\mathrm{B}/dV_\mathrm{A}$ were extracted. Since we are interested in non-local conductances which are smaller than local conductances by orders of magnitude, care was taken to ensure that the measured signals were not affected by phase shifts or crosstalk between measurement lines. We performed extensive electronic circuit simulations of the entire measurement setup including amplifiers, filters, and cryostat wiring. The simulations showed that the signals are reliable for measurement frequencies up to $f\approx 200~\mathrm{Hz}$, which was also confirmed experimentally. All data shown in this paper were taken at $f\approx 37~\mathrm{Hz}$ with a typical AC amplitude of $5~\mathrm{\mu V}$. In the remainder of the paper, $V_\mathrm{A}$ and $V_\mathrm{B}$ will refer to DC bias voltage, $G$ will refer to normal-state conductance, and $g$ will refer to differential conductance.

\begin{figure}
\includegraphics[width=\columnwidth]{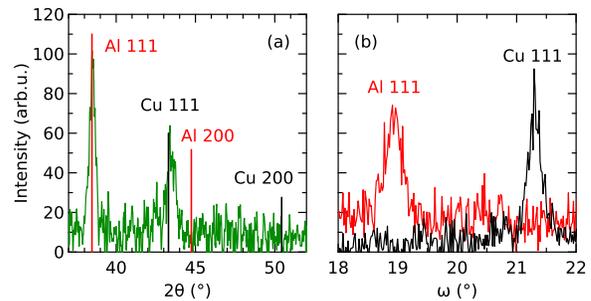}
\caption{\label{fig_xray}(Color online) X-ray diffraction patterns of an unpatterend Al/oxide/Cu sandwich, with substrate background subtracted. (a) $\theta-2\theta$ scan, with expected positions and relative intensities of (111) and (200) reflections for both Al and Cu indicated by vertical bars. (b) Rocking curve with angle of inclination $\omega$ for the (111) reflections of Al and Cu.}
\end{figure}

For X-ray characterization, we also prepared unpatterned bilayers in the same way as the overlap contact area of the patterned samples, i.e., with a first layer of aluminum, subsequent oxidation, and a second copper layer on top. X-ray diffraction of these bilayer films was performed using copper K$\alpha$ radiation and a solid-state detector. Figure \ref{fig_xray}(a) shows $\theta-2\theta$ scans, with substrate background subtracted. The only reflections we observed were the (111) reflections for both Al and Cu. As an example, the expected positions of the (200) reflections, and their expected intensities relative to the (111) reflections, are also indicated by vertical bars for both materials. In Fig.~\ref{fig_xray}(b), rocking curves for the two (111) reflections are shown, with an FWHM of $0.35^\circ$ and $0.26^\circ$ for Al and Cu, respectively. From the absence of all but the (111) reflections, and their small width in the rocking curve, we conclude that the films have a nearly perfect (111) texture.

\section{Results}

\begin{figure}
\includegraphics[width=\columnwidth]{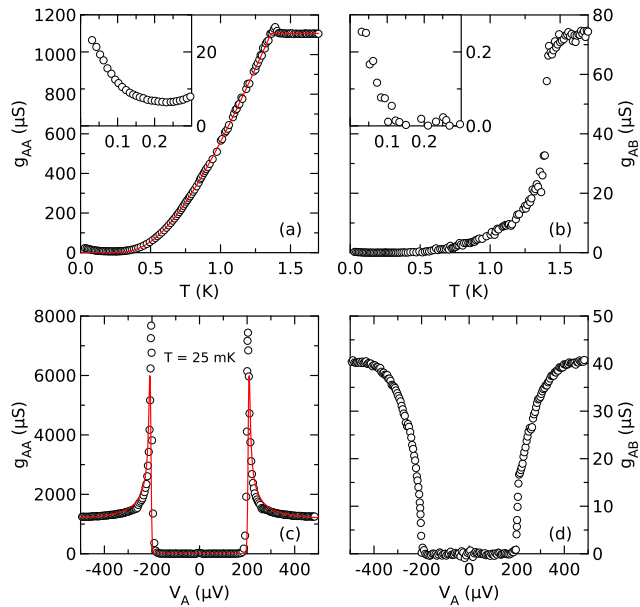}
\caption{\label{fig_gfull}(Color online)
Local conductance $g_\mathrm{AA}$ (a,c) and non-local conductance $g_\mathrm{AB}$ (b,d) of sample I as a function of temperature $T$ at $V_\mathrm{A}=0$ (a,b) and injector bias $V_\mathrm{A}$ at $T=25~\mathrm{mK}$ (c,d). The insets in (a) and (b) show the low-temperature region of both plots on an enlarged scale. The solid lines in (a) and (c) are fits to a standard BCS tunneling model.}
\end{figure}

The local conductance $g_\mathrm{AA}$ of the injector contact A of sample I is shown in Fig.~\ref{fig_gfull}(a) as a function of temperature $T$ for $V_\mathrm{A}=0$. Above the critical temperature $T_\mathrm{c}=1.35~\mathrm{K}$ of the aluminum film, a constant normal-state tunnel conductance is observed. Below $T_\mathrm{c}$, the conductance gradually drops to zero, and can be fitted with a standard BCS\cite{bardeen1957,giaever1960} tunneling model, with zero-temperature gap $\Delta_0=205~\mathrm{\mu eV}$ and normal-state conductance $G_\mathrm{A}=1105~\mathrm{\mu S}$. Below about 250~mK, an additional conductance contribution not covered by the tunneling model is observed. This anomaly can be better resolved in the inset, where it is shown on an enlarged scale. Figure \ref{fig_gfull}(b) shows the non-local conductance $g_\mathrm{AB}$ measured simultaneously. The overall behavior is similar to the local conductance, including the presence of a low-temperature anomaly, as seen in the inset. The low-temperature anomaly decreases more steeply with increasing temperature as compared to the local anomaly seen in the inset of Fig.~\ref{fig_gfull}(a), as will be discussed later.

The local differential conductance $g_\mathrm{AA}$ for the same configuration is shown in Fig.~\ref{fig_gfull}(c) as a function of bias $V_\mathrm{A}$ at $T=20~\mathrm{mK}$. It has the form of the BCS density of states as expected for a tunnel contact with low transparency ($\mathcal{T}\approx 6 \cdot 10^{-5}$ for this sample, estimated from $G_\mathrm{A}$ and the contact area). The solid line is a fit to the BCS tunneling model, with the same parameters as in Fig.~\ref{fig_gfull}(a). Except for the small zero-bias, low-temperature anomaly already seen in panel (a), and discussed further below, the subgap conductance is negligible, indicating a high quality tunnel barrier without pin holes. The non-local differential conductance $g_\mathrm{AB}$ corresponding to panel (c) is shown in panel (d). Similar to the local signal, the subgap conductance is almost zero, with a sharp transition to a finite signal above the gap. The signal above the gap can be attributed to charge imbalance caused by quasiparticle injection. A detailed investigation of the non-local charge-imbalance signal will be reported elsewhere.

\begin{figure}
\includegraphics[width=\columnwidth]{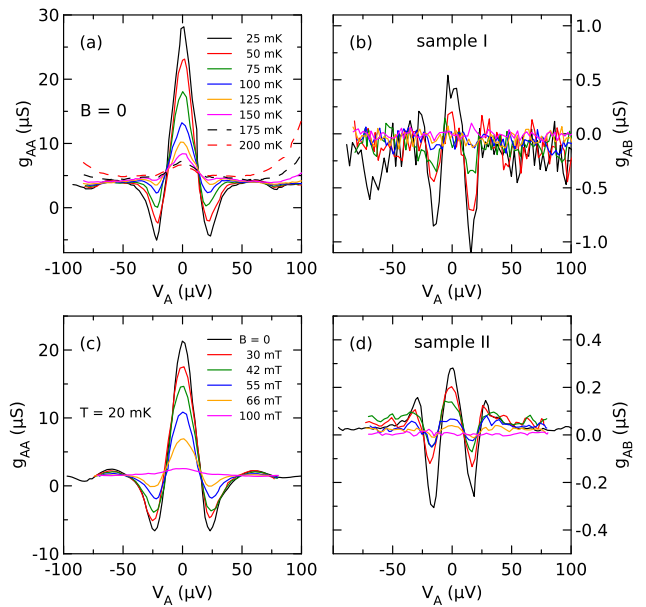}
\caption{\label{fig_tbdep}(Color online)
Local differential conductance $g_\mathrm{AA}$ (a,c) and non-local differential conductance $g_\mathrm{AB}$ (b,d) as a function of injector bias $V_\mathrm{A}$ taken at different temperatures $T$ for sample I (a,b), and different magnetic fields $B$ for sample II (c,d).}
\end{figure}

Figure \ref{fig_tbdep} shows the low-temperature subgap anomalies seen in both local and non-local conductance on an enlarged scale. Panel (a) shows the local differential conductance of sample I for different temperatures in the range between 25~mK and 200~mK, at zero magnetic field. A sharp peak is observed at zero bias. The peak height decreases with increasing temperature, as already observed in the inset of Fig.~\ref{fig_gfull}(a). At $V_\mathrm{A}\approx \pm 20~\mathrm{\mu V}$, minima are seen, and for the lowest temperatures the differential conductance actually becomes negative. At higher bias, there are a series of side-maxima, which will be discussed in more detail below (see Fig.~\ref{fig_fits}). The positions of the maxima and minima are independent of temperature. Panel (b) shows the non-local differential conductance measured simultaneously. The non-local conductance shows a structure similar to the local one, with a central peak and negative side-minima. The minima occur at $V_\mathrm{A}\approx \pm 15~\mathrm{\mu V}$, sligthly below the position of the minima in the local conductance. The anomaly decreases with temperature, and drops below the noise floor above 100~mK. Panels (c) and (d) show the local and non-local differential conductances of sample II at low temperature for different magnetic fields $B$ applied parallel to the substrate plane along the direction of the copper wires forming the contacts. The signals are almost identical to those observed in sample I. As temperature, the magnetic field leads to a suppression of the anomalies, on a field scale much below the critical field of the aluminum wire ($B_\mathrm{c}\approx 600~\mathrm{mT}$, determined from conductance measurements). While the amplitude of the signals is suppressed by both finite temperature and magnetic field, no change in the overall shape or voltage scales is seen.

\begin{figure}
\includegraphics[width=\columnwidth]{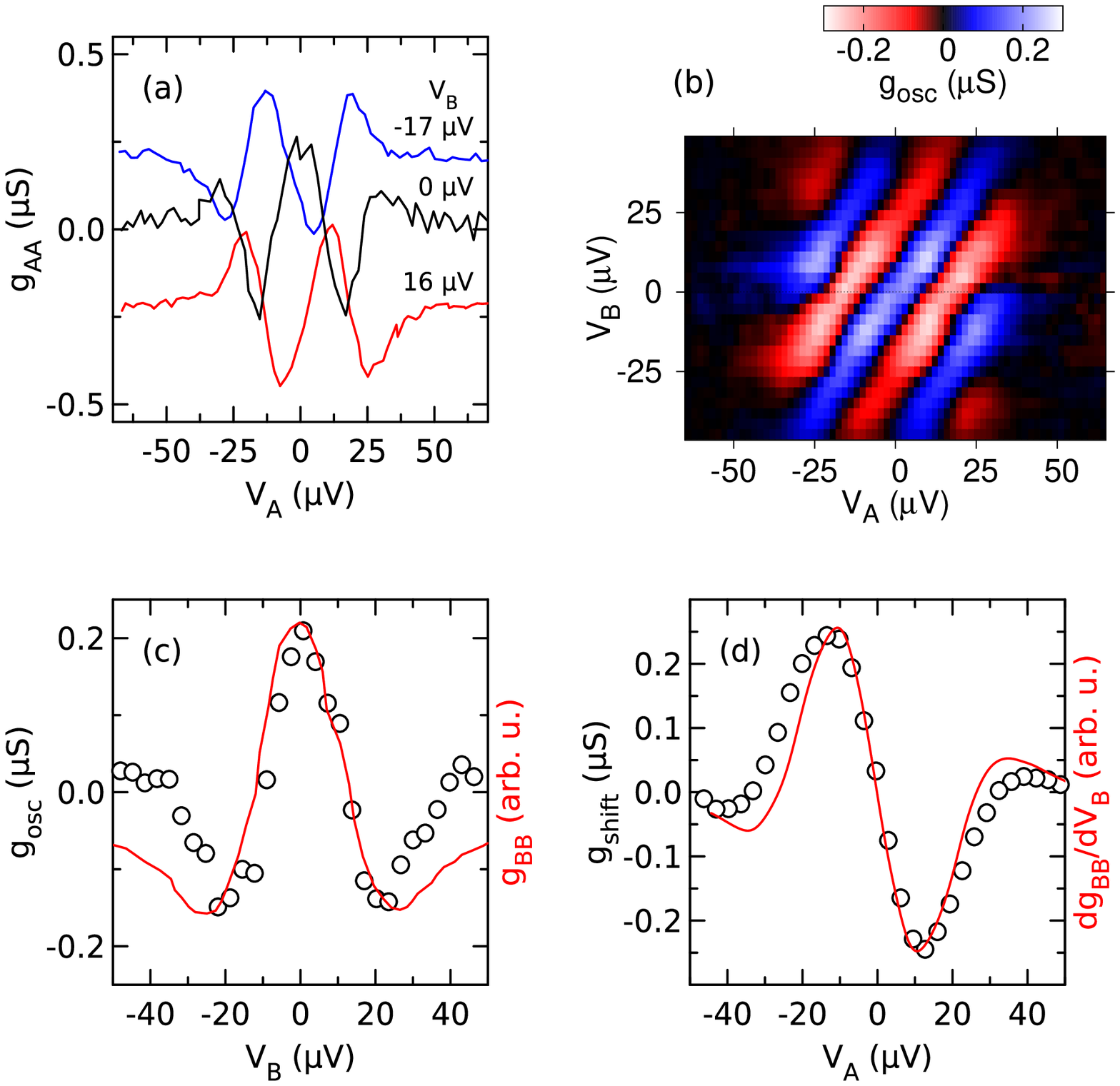}
\caption{\label{fig_nlmap}(Color online)\\
(a) $g_\mathrm{AB}$ as a function of $V_\mathrm{A}$ for different $V_\mathrm{B}$.\\
(b) Oscillatory part $g_\mathrm{osc}$ of the non-local conductance $g_\mathrm{AB}$ as a function of both $V_\mathrm{A}$ and $V_\mathrm{B}$.\\
(c) $g_\mathrm{osc}$ at $V_\mathrm{A}=0$ (symbols) and local conductance $g_\mathrm{BB}$ of contact B (line) as a function of $V_\mathrm{B}$.\\
(d) $g_\mathrm{shift}$ (symbols) and derivative of the local conductance $dg_\mathrm{BB}/dV_\mathrm{B}$ as a function of $V_\mathrm{B}$ (line).\\
All data are taken at $B=0$ and $T=20~\mathrm{mK}$ from sample II. For an explanation of $g_\mathrm{osc}$ and $g_\mathrm{shift}$, see text.}
\end{figure}

To study the non-local conductance pattern in more detail, we have applied an additional DC bias $V_\mathrm{B}$ to the second contact. Figure \ref{fig_nlmap}(a) shows $g_\mathrm{AB}$ as a function of $V_\mathrm{A}$ for three different values of $V_\mathrm{B}$. The effect of $V_\mathrm{B}$ is twofold: first, there is an overall vertical shift independent of $V_\mathrm{A}$, as can be clearly seen at large bias. Second, there is a horizontal shift of the maxima and minima of the oscillatory pattern. We can therefore describe the signal as

\[g_\mathrm{AB}(V_\mathrm{A},V_\mathrm{B})=g_\mathrm{shift}(V_\mathrm{B})+g_\mathrm{osc}(V_\mathrm{A},V_\mathrm{B})\]

The vertical shift $g_\mathrm{shift}$ can be extracted from the data at bias voltages around $60~\mathrm{\mu V}$, where the oscillatory signal $g_\mathrm{osc}$ has died out. $g_\mathrm{osc}$ obtained by subtracting $g_\mathrm{shift}$ is shown in Fig.~\ref{fig_nlmap}(b) in a color plot as a function of both $V_\mathrm{A}$ and $V_\mathrm{B}$. The pattern is limited to a bias voltage window $|V_\mathrm{A},V_\mathrm{B}|\leq 40~\mathrm{\mu V}$, and the shift of the maxima and minima is linear in $V_\mathrm{B}$, with a slope of about 0.7 (i.e., the signal does not simply depend on $V_\mathrm{A}-V_\mathrm{B}$). It can also be seen that additional maxima and minima shift into the bias voltage window upon increasing $|V_\mathrm{B}|$.

Figure \ref{fig_nlmap}(c) shows $g_\mathrm{osc}$ as a function of $V_\mathrm{B}$ for $V_\mathrm{A}=0$ (i.e., a vertical cut through Fig.~\ref{fig_nlmap}(b), together with the local conductance $g_\mathrm{BB}$ of contact B, measured independently. As can be seen, the two signals roughly scale onto each other. A further correlation between non-local and local data can be seen in Fig.~\ref{fig_nlmap}(d), where $g_\mathrm{shift}$ is plotted as a function of $V_\mathrm{B}$, along with the second derivative $dg_\mathrm{BB}/dV_\mathrm{B}=d^2I_\mathrm{B}/dV^2_\mathrm{B}$ of the local conductance of contact B.

\begin{figure}
\includegraphics[width=\columnwidth]{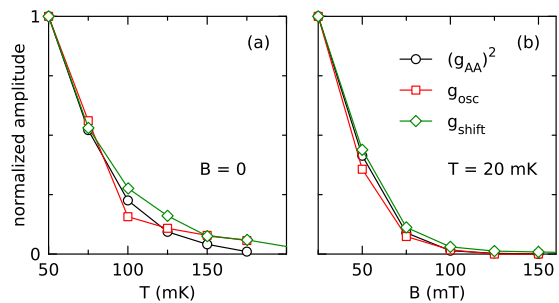}
\caption{\label{fig_amp}(Color online)
Normalized amplitudes of $g_\mathrm{osc}$ and $g_\mathrm{shift}$, together with the square of the amplitude of the subgap anomaly in the local conductance $(g_\mathrm{AA})^2$, as a function of  temperature $T$ (a) and applied magnetic field $B$ (b).\\
Data taken from sample I.}
\end{figure}

Figure~\ref{fig_amp}(a) and (b) show the dependence of the amplitudes of the local and non-local subgap anomalies as a function of temperature and magnetic field. Data are taken from sample I, and normalized to the low-temperature and low-field values. Both the local and non-local signals decay rapidly, at temperature and field scales well below the critical temperature $T_\mathrm{c}=1.3~\mathrm{K}$ and critical field $B_\mathrm{c}=600~\mathrm{mT}$, respectively. As can be seen, the two components $g_\mathrm{osc}$ and $g_\mathrm{shift}$ of the non-local conductance both scale with the square of the amplitude of the zero-bias anomaly in the local conductance. The steeper decrease of the non-local conductance as a function of temperature has already been noted in Fig.~\ref{fig_gfull}.

\begin{figure}
\includegraphics[width=\columnwidth]{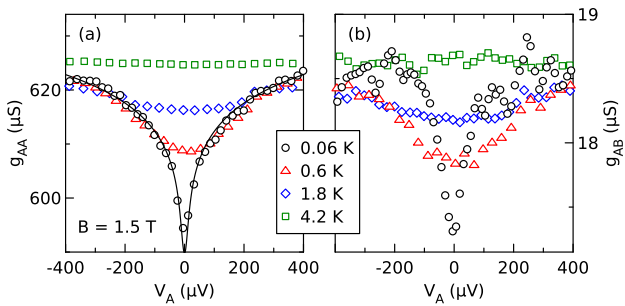}
\caption{\label{fig_dcb}(Color online)
Local conductance $g_\mathrm{AA}$ (a) and non-local conductance $g_\mathrm{AB}$ (b) of sample II in the normal state at $B=1.5~\mathrm{T}$ for different temperatures $T$. The solid line in panel (a) is a fit to the standard model of dynamical Coulomb blockade.\cite{devoret1990}}
\end{figure}

In order to elucidate the role of dynamical Coulomb blockade (DCB) in our samples, we have also investigated the differential conductance in the normal state, where the impact of DCB is well known. Figure \ref{fig_dcb} shows both local (a) and non-local (b) differential conductance in a magnetic field of $1.5~\mathrm{T}$, sufficiently large to suppress superconductivity, at different temperatures. At lowest temperature, the local conductance shows a low-bias dip which can be well described by DCB in the presence of an Ohmic environment,\cite{devoret1990} with environmental impedance $R_\mathrm{env}=160~\mathrm{\Omega}$, see fit in Fig.~\ref{fig_dcb}(a). The Coulomb dip persists up to temperatures of about 2~K, well above the critical temperature of aluminum. A similar dip appears in the non-local conductance, shown in panel (b). The dip in the non-local conductance is slightly narrower than in the local conductance, and also persists up to about 2~K. 

\section{Discussion}

\subsection{Local conductance}

\begin{figure}
\includegraphics[width=\columnwidth]{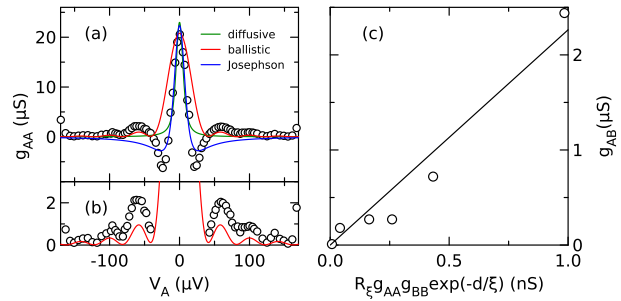}
\caption{\label{fig_fits}(Color online) (a,b) Differential subgap conductance of sample II at $B=0$ and $T=20~\mathrm{mK}$, with regular background subtracted (symbols). The lines are three different fits described in detail in the text. (c) Measured non-local conductance amplitude $g_\mathrm{AB}$ vs. model prediction $g_\mathrm{AB}= R_\xi\cdot g_\mathrm{AA}\cdot g_\mathrm{BB}\cdot \exp(-d/\xi)$. See text for details.}
\end{figure}

We will first focus on the interpretation of the subgap anomaly of the local conductance. Figure \ref{fig_fits}(a) shows the differential conductance $g_\mathrm{AA}$ of sample II at low temperature and zero field, together with three different models described in the following. Here, a small regular background contribution of about $1~\mathrm{\mu S}$ has been subtracted to show the anomaly more clearly. Panel (b) shows the same data on an enlarged scale, together with the ballistic model.

{\em Diffusive model}: A zero-bias conductance peak is known to occur in low-transparency tunnel junctions between diffusive normal metals and superconductors as a result of phase-coherent enhancement of Andreev reflection (reflectionless tunneling).\cite{kastalsky1991,vanwees1992,beenakker1992} The subgap conductance due to reflectionless tunneling for different experimental situations has been calculated in Refs. \onlinecite{volkov1992,volkov1993,volkov1993b,volkov1994}. A fit to the model of an overlap junction including pair-breaking (equation (6) of Ref.~\onlinecite{volkov1994}) shows good agreement for the central peak, but fails to describe the addtional minima and maxima. It should be noted that the characteristic energy scale $\epsilon_\mathrm{N}=0.18~\mathrm{\mu eV}$ controlling the weight of the anomaly was actually not fitted but simply calculated from known sample parameters. Only a small amount of pair-breaking ($\gamma_\mathrm{N}=4~\mathrm{\mu eV}$) had to be adjusted to obtain the correct width (see Ref.~\onlinecite{volkov1994} for the definition of the parameters).

{\em Ballistic model}: Reflectionless tunneling results from the constructive interference of repeated attempts of an electron to be Andreev-reflected at the tunnel junction. In a thin-film overlap junction these repeated attempts will mostly come from electrons bouncing back and forth between the junction and the upper surface of the normal metal film. Elastic scattering in thin films can largely be attributed to grain boundaries and surfaces, since both the elastic mean free path and the grain size are usually of the order of the film thickness. Therefore, while electron motion in the lateral direction is certainly diffusive, it can be assumed to be close to ballistic perpendicular to the films, especially since our films show a nearly perfect (111) texture. The ballistic limit of reflectionless tunneling has been considered in Ref.~\onlinecite{schechter2001} using semiclassical trajectories, similar to the treatment of diffusive motion in Ref.~\onlinecite{vanwees1992}. Including the phase difference due to finite electron energy $E$, $\Delta\phi=2EL/\hbar v_\mathrm{F}$,\cite{vanwees1992} where $L$ is the length of a trajectory and $v_\mathrm{F}$ is the Fermi velocity, we obtain a differential conductance spectrum with a central peak and small, equidistant side-maxima, as seen best on a larger scale in Fig.~\ref{fig_fits}(b). The fit parameters were chosen such that the height of the central peak and the position of the side-maxima are reproduced. Better fits for either the central peak or the side-maxima could be obtained, but only at the expense of the other. In addition to the position of the side-maxima, also their decay at higher bias is reproduced. The negative differential conductance remains unexplained in this model.

{\em Josephson model}: A zero-bias peak with a subsequent region of negative differential conductance can be observed in voltage biased Josephson junctions in the phase-diffusion regime. A part of the normal-metal leads of our junctions, the green shaded region in Fig.~\ref{fig_sem}(b), actually consists of a Cu/Al/oxide/Cu trilayer as a result of shadow evaporation. While superconductivity in the aluminum of this trilayer must be strongly suppressed by the inverse proximity effect, we can not exclude the presence of a weak Josephson coupling between the trilayer and the aluminum bar. As in the case of reflectionless tunneling, the bottleneck for the Josephson coupling would be pair transmission through the tunnel barrier. The model of phase diffusion\cite{ingold1994} plotted in Fig.~\ref{fig_fits}(a) gives a qualitative description of the central peak and negative differential conductance, but no side-maxima are predicted here. The phase diffusion model predicts that the current maximum (i.e., the zero of the differential conductance) appears at a voltage which is proportional to temperature. In contrast, no temperature dependence of the voltage scales is observed in our samples. In the light of the Josephson model, one may ask whether the additional side-maxima are the result of multiple Andreev reflection (MAR).\cite{octavio1983} Peaks due to MAR would appear at characteristic voltages $V=2\Delta/ne$, where $n$ is an integer corresponding to the correlated transfer of $n$ electrons. MARs are inconsistent with our data for several reasons: first, they are observable only at relatively high contact transparency due to the multiple transmission through the interface. For the same reason, the features with smallest $n$, i.e., at highest bias, should be the largest, in contrast to our observations. Also, the observed side-maxima are equidistant, and too few for MAR (six should be visible between $50$ and $150~\mathrm{\mu eV}$, assuming the bulk gap $\Delta=200~\mathrm{\mu eV}$). 

All three models have in common that they require phase-coherent motion of Andreev pairs on the normal metal side of the tunnel junction. This is consistent with the fast suppression of the conductance anomaly both as a function of temperature, and in the presence of a magnetic field. We conclude that the oscillatory structure in the local conductance is caused by reflectionless tunneling close to the ballistic limit, while the negative differential conductance might indicate the presence of a weak Josephson coupling.

Finally, we would like to address the impact of dynamical Coulomb blockade (DCB) in the presence of a finite-impedance electromagnetic environment on the local conductance. As already discussed (see Fig.~\ref{fig_dcb}), in the normal state the local conductance exhibits a suppression at low bias, which can be explained by DCB with an Ohmic environment.\cite{devoret1990} Considering our sample design, we expect the series impedance to be mainly given by the long, narrow aluminum wire. Indeed, the resistance of the aluminum wire between tunnel junction B and lead 1 (see Fig.~\ref{fig_sem}) is $190~\mathrm{\Omega}$, similar to $R_\mathrm{env}=160~\mathrm{\Omega}$ extracted from the fit. In the superconducting state, Coulomb blockade of subgap transport is expected to be stronger than in the normal state due to the double charge transfer of Andreev reflection. Despite the presence of dynamical Coulomb blockade, we observe a phase-coherent enhancement of subgap conductance in the superconducting state.

\subsection{Non-local conductance}

Theoretical descriptions of CAR and EC in lowest order of contact transparency, i.e., in the tunneling limit, predict the probability of both processes to be proportional to 
$G_\mathrm{A}G_\mathrm{B}\zeta$, where $G_\mathrm{A}$ and $G_\mathrm{B}$ are the normal-state conductances of the two contacts, and $\zeta$ is a factor related to the propagation of virtual quasiparticles between the contacts in the superconductor. $\zeta$ depends on sample geometry, the mean free path and the coherence length $\xi$.\cite{falci2001,feinberg2003} In a one-dimensional geometry with two point contacts at a distance $d$, $\zeta=\exp(-d/\xi)$. Since CAR and EC contribute with opposite signs to the non-local conductance, their effect is expected to cancel in the tunneling limit, i.e., $g_\mathrm{AB}=0$. This is in contrast to our observation of a positive non-local conductance (corresponding to EC) at low bias, followed by a cross-over to a negative signal (corresponding to CAR) at higher bias. As can be seen in Fig.~\ref{fig_nlmap}(b), the crossover is actually part of a regular oscillatory pattern as a function of both injector and detector bias voltage. We have found several correlations between the local and non-local subgap anomalies: Both exhibit a central peak and an oscillatory dependence on bias voltage. Some aspects of the local and non-local signals can be scaled onto each other (Figs.~\ref{fig_nlmap}c and d). In particular, the amplitude of the non-local conductance anomaly scales with the square of the local one, both as a function of temperature and magnetic field. This shows conclusively that both must have a common origin.

It has been predicted recently that reflectionless tunneling in diffusive NSN structures leads to a subgap anomaly in the non-local conductance which is proportional to the product of the subgap anomalies in the local conductance of injector and detector contact, i.e., $g_\mathrm{AB}\propto g_\mathrm{AA}g_\mathrm{BB}$.\cite{golubev2009} This is similar to the prediction of the lowest-order tunneling model discussed above, except that the {\em normal-state conductances} $G_\mathrm{A}$ and $G_\mathrm{B}$ are replaced by the {\em actual subgap anomalies} $g_\mathrm{AA}$ and $g_\mathrm{BB}$. Since the model in Ref.~\onlinecite{golubev2009} assumes purely diffusive motion, it only predicts a zero-bias peak, but no additional oscillatory structure. For one-dimensional systems, the quantitative prediction is $g_\mathrm{AB}\approx R_\xi\cdot g_\mathrm{AA}\cdot g_\mathrm{BB}\cdot\exp(-d/\xi)$, where $R_\xi$ is the normal-state resistance of the superconducting wire on a length of $\xi$. We cannot reproduce the non-local conductance pattern in detail with this model, and will therefore restrict ourselves to a discussion of the signal amplitudes. First we note that the scaling of $g_\mathrm{AB}$ with $g_\mathrm{AA}^2$ seen in Fig.~\ref{fig_amp} is fully consistent with the model, since within one sample $g_\mathrm{AA}\approx g_\mathrm{BB}$. For a comparison of the different samples, we have plotted the measured non-local conductance amplitude against the prediction in Fig.~\ref{fig_fits}(c). As can be seen, the scaling is obeyed reasonably well, but the observed non-local signals are too large by three orders of magnitude. The scaling is expected to hold for arbitrary sample geometries, while the quantitative prediction is for the specific case of one-dimensional systems. In the one-dimensional case, the same model predicts the local anomaly to be $g_\mathrm{AA}\approx R_\mathrm{N}G_\mathrm{A}^2$, where $R_\mathrm{N}$ is the resistance of the copper wire over the normal-metal coherence length $\xi_\mathrm{N}=\sqrt{\hbar D/\epsilon}$. For our sample parameters, this underestimates the local anomaly by about one order of magnitude, whereas the model of an extended overlap junction used in Fig.~\ref{fig_fits}(a) provides the correct signal amplitude. Whether a theoretical model of our specific non-local geometry would similarly remedy the quantitative disagreement seen in Fig.~\ref{fig_fits}(c) remains an open question. Also, a model treating the impact of reflectionless tunneling on non-local transport in ballistic structures would be highly desirable to see whether the oscillatory structure seen in the non-local conductance can be reproduced.

An alternative explanation for the finite non-local subgap conductance is dynamical Coulomb blockade. In Ref.~\onlinecite{yeyati2007}, a non-local version of DCB has been predicted to lift the exact cancellation of CAR and EC in the tunneling limit. Depending on whether the blockade of CAR or EC is stronger, the other process dominates, and $g_\mathrm{AB}$ may be either positive or negative. In this model, the blockade of CAR and EC is controlled by coupling to electromagnetic modes of different symmetry. Since these modes, in general, have different energies, a bias-dependent crossover from dominating EC to CAR can be explained. The situation is simpler in the normal state, where only EC is possible. In this case, the expectation is a suppression of both local and non-local transport at low bias, as indeed observed in Fig.~\ref{fig_dcb}. However, the Coulomb dip in both local and non-local conductance persists to temperatures $T>1~\mathrm{K}$, which is inconsistent with the temperature range $T\leq 150~\mathrm{mK}$ of the subgap anomalies in the superconducting state. Also, DCB is not affected by a magnetic field (the data in Fig.~\ref{fig_dcb} were taken at $B=1.5~\mathrm{T}$), whereas the subgap anomalies are restricted to $B\leq 100~\mathrm{mT}$. The differences in energy, temperature, and magnetic field scales clearly show that Coulomb interaction, while present in our samples, is not the cause of the subgap anomalies in the superconducting state.

\section{Conclusion}

In conclusion we have presented experimental results on non-local transport in mesoscopic normal-metal/superconductor hybrid structures. Local and non-local conductance exhibit correlated subgap anomalies, which are shown to be due to phase-coherent enhancement of transport at low energies. A systematic dependence of non-local transport on bias conditions is observed, which allows control over the dominating transport processes. Dynamical Coulomb blockade can be ruled out as the cause of the subgap anomalies. A comprehensive theoretical description is highly desirable as a guideline for the design of superconducting solid-state entanglers.

We acknowledge W. Belzig, M. Eschrig and C. S\"urgers for useful discussions. This work was supported by the Landesstiftung Baden-W\"urttemberg within the "Kompetenznetz Funktionelle Nanostrukturen", and by the DFG Center for Functional Nanostructures.

\bibliography{../../../lit.bib}

\begin{thebibliography}{34}
\expandafter\ifx\csname natexlab\endcsname\relax\def\natexlab#1{#1}\fi
\expandafter\ifx\csname bibnamefont\endcsname\relax
  \def\bibnamefont#1{#1}\fi
\expandafter\ifx\csname bibfnamefont\endcsname\relax
  \def\bibfnamefont#1{#1}\fi
\expandafter\ifx\csname citenamefont\endcsname\relax
  \def\citenamefont#1{#1}\fi
\expandafter\ifx\csname url\endcsname\relax
  \def\url#1{\texttt{#1}}\fi
\expandafter\ifx\csname urlprefix\endcsname\relax\def\urlprefix{URL }\fi
\providecommand{\bibinfo}[2]{#2}
\providecommand{\eprint}[2][]{\url{#2}}

\bibitem[{\citenamefont{Andreev}(1964)}]{andreev1964}
\bibinfo{author}{\bibfnamefont{A.~F.} \bibnamefont{Andreev}},
  \bibinfo{journal}{Sov. Phys. JETP} \textbf{\bibinfo{volume}{19}},
  \bibinfo{pages}{1228} (\bibinfo{year}{1964}).

\bibitem[{\citenamefont{Byers and Flatt\'e}(1995)}]{byers1995}
\bibinfo{author}{\bibfnamefont{J.~M.} \bibnamefont{Byers}} \bibnamefont{and}
  \bibinfo{author}{\bibfnamefont{M.~E.} \bibnamefont{Flatt\'e}},
  \bibinfo{journal}{Phys. Rev. Lett.} \textbf{\bibinfo{volume}{74}},
  \bibinfo{pages}{306} (\bibinfo{year}{1995}).

\bibitem[{\citenamefont{Deutscher and Feinberg}(2000)}]{deutscher2000}
\bibinfo{author}{\bibfnamefont{G.}~\bibnamefont{Deutscher}} \bibnamefont{and}
  \bibinfo{author}{\bibfnamefont{D.}~\bibnamefont{Feinberg}},
  \bibinfo{journal}{Appl. Phys. Lett.} \textbf{\bibinfo{volume}{76}},
  \bibinfo{pages}{487} (\bibinfo{year}{2000}).

\bibitem[{\citenamefont{Falci et~al.}(2001)\citenamefont{Falci, Feinberg, and
  Hekking}}]{falci2001}
\bibinfo{author}{\bibfnamefont{G.}~\bibnamefont{Falci}},
  \bibinfo{author}{\bibfnamefont{D.}~\bibnamefont{Feinberg}}, \bibnamefont{and}
  \bibinfo{author}{\bibfnamefont{F.~W.~J.} \bibnamefont{Hekking}},
  \bibinfo{journal}{Europhys. Lett.} \textbf{\bibinfo{volume}{54}},
  \bibinfo{pages}{255} (\bibinfo{year}{2001}).

\bibitem[{\citenamefont{Beckmann et~al.}(2004)\citenamefont{Beckmann, Weber,
  and L\"ohneysen}}]{beckmann2004}
\bibinfo{author}{\bibfnamefont{D.}~\bibnamefont{Beckmann}},
  \bibinfo{author}{\bibfnamefont{H.~B.} \bibnamefont{Weber}}, \bibnamefont{and}
  \bibinfo{author}{\bibfnamefont{H.~v.} \bibnamefont{L\"ohneysen}},
  \bibinfo{journal}{Phys. Rev. Lett.} \textbf{\bibinfo{volume}{93}},
  \bibinfo{pages}{197003} (\bibinfo{year}{2004}).

\bibitem[{\citenamefont{Beckmann and L\"ohneysen}(2007)}]{beckmann2007}
\bibinfo{author}{\bibfnamefont{D.}~\bibnamefont{Beckmann}} \bibnamefont{and}
  \bibinfo{author}{\bibfnamefont{H.~v.} \bibnamefont{L\"ohneysen}},
  \bibinfo{journal}{Appl. Phys. A} \textbf{\bibinfo{volume}{89}},
  \bibinfo{pages}{603} (\bibinfo{year}{2007}).

\bibitem[{\citenamefont{Russo et~al.}(2005)\citenamefont{Russo, Kroug,
  Klapwijk, and Morpurgo}}]{russo2005}
\bibinfo{author}{\bibfnamefont{S.}~\bibnamefont{Russo}},
  \bibinfo{author}{\bibfnamefont{M.}~\bibnamefont{Kroug}},
  \bibinfo{author}{\bibfnamefont{T.~M.} \bibnamefont{Klapwijk}},
  \bibnamefont{and} \bibinfo{author}{\bibfnamefont{A.~F.}
  \bibnamefont{Morpurgo}}, \bibinfo{journal}{Phys. Rev. Lett.}
  \textbf{\bibinfo{volume}{95}}, \bibinfo{pages}{027002}
  (\bibinfo{year}{2005}).

\bibitem[{\citenamefont{Cadden-Zimansky and Chandrasekhar}(2006)}]{cadden2006}
\bibinfo{author}{\bibfnamefont{P.}~\bibnamefont{Cadden-Zimansky}}
  \bibnamefont{and}
  \bibinfo{author}{\bibfnamefont{V.}~\bibnamefont{Chandrasekhar}},
  \bibinfo{journal}{Phys. Rev. Lett.} \textbf{\bibinfo{volume}{97}},
  \bibinfo{pages}{237003} (\bibinfo{year}{2006}).

\bibitem[{\citenamefont{Cadden-Zimansky
  et~al.}(2007)\citenamefont{Cadden-Zimansky, Jiang, and
  Chandrasekhar}}]{cadden2007}
\bibinfo{author}{\bibfnamefont{P.}~\bibnamefont{Cadden-Zimansky}},
  \bibinfo{author}{\bibfnamefont{Z.}~\bibnamefont{Jiang}}, \bibnamefont{and}
  \bibinfo{author}{\bibfnamefont{V.}~\bibnamefont{Chandrasekhar}},
  \bibinfo{journal}{New J. Phys.} \textbf{\bibinfo{volume}{9}},
  \bibinfo{pages}{116} (\bibinfo{year}{2007}).

\bibitem[{\citenamefont{Cadden-Zimansky
  et~al.}(2009)\citenamefont{Cadden-Zimansky, Wei, and
  Chandrasekhar}}]{cadden2009}
\bibinfo{author}{\bibfnamefont{P.}~\bibnamefont{Cadden-Zimansky}},
  \bibinfo{author}{\bibfnamefont{J.}~\bibnamefont{Wei}}, \bibnamefont{and}
  \bibinfo{author}{\bibfnamefont{V.}~\bibnamefont{Chandrasekhar}},
  \bibinfo{journal}{Nature Physics} \textbf{\bibinfo{volume}{5}},
  \bibinfo{pages}{393} (\bibinfo{year}{2009}).

\bibitem[{\citenamefont{Asulin et~al.}(2006)\citenamefont{Asulin, Yuli, Koren,
  and Millo}}]{asulin2006}
\bibinfo{author}{\bibfnamefont{I.}~\bibnamefont{Asulin}},
  \bibinfo{author}{\bibfnamefont{O.}~\bibnamefont{Yuli}},
  \bibinfo{author}{\bibfnamefont{G.}~\bibnamefont{Koren}}, \bibnamefont{and}
  \bibinfo{author}{\bibfnamefont{O.}~\bibnamefont{Millo}},
  \bibinfo{journal}{Phys. Rev. B} \textbf{\bibinfo{volume}{74}},
  \bibinfo{pages}{092501} (\bibinfo{year}{2006}).

\bibitem[{\citenamefont{Kleine et~al.}(2009)\citenamefont{Kleine, Baumgartner,
  Trbovic, and Sch\"onenberger}}]{kleine2009}
\bibinfo{author}{\bibfnamefont{A.}~\bibnamefont{Kleine}},
  \bibinfo{author}{\bibfnamefont{A.}~\bibnamefont{Baumgartner}},
  \bibinfo{author}{\bibfnamefont{J.}~\bibnamefont{Trbovic}}, \bibnamefont{and}
  \bibinfo{author}{\bibfnamefont{C.}~\bibnamefont{Sch\"onenberger}},
  \bibinfo{journal}{Europhys. Lett.} \textbf{\bibinfo{volume}{87}},
  \bibinfo{pages}{27011} (\bibinfo{year}{2009}).

\bibitem[{\citenamefont{Hofstetter et~al.}(2009)\citenamefont{Hofstetter,
  Csonka, Nyg\aa{}rd, and Sch\"onenberger}}]{hofstetter2009}
\bibinfo{author}{\bibfnamefont{L.}~\bibnamefont{Hofstetter}},
  \bibinfo{author}{\bibfnamefont{S.}~\bibnamefont{Csonka}},
  \bibinfo{author}{\bibfnamefont{J.}~\bibnamefont{Nyg\aa{}rd}},
  \bibnamefont{and}
  \bibinfo{author}{\bibfnamefont{C.}~\bibnamefont{Sch\"onenberger}},
  \bibinfo{journal}{Nature} \textbf{\bibinfo{volume}{461}},
  \bibinfo{pages}{960} (\bibinfo{year}{2009}).

\bibitem[{\citenamefont{Burkard}(2007)}]{burkard2007}
\bibinfo{author}{\bibfnamefont{G.}~\bibnamefont{Burkard}}, \bibinfo{journal}{J.
  Phys.: Condens. Matter} \textbf{\bibinfo{volume}{19}},
  \bibinfo{pages}{233202} (\bibinfo{year}{2007}).

\bibitem[{\citenamefont{Morten et~al.}(2006)\citenamefont{Morten, Brataas, and
  Belzig}}]{morten2006}
\bibinfo{author}{\bibfnamefont{J.~P.} \bibnamefont{Morten}},
  \bibinfo{author}{\bibfnamefont{A.}~\bibnamefont{Brataas}}, \bibnamefont{and}
  \bibinfo{author}{\bibfnamefont{W.}~\bibnamefont{Belzig}},
  \bibinfo{journal}{Phys. Rev. B} \textbf{\bibinfo{volume}{74}},
  \bibinfo{pages}{214510} (\bibinfo{year}{2006}).

\bibitem[{\citenamefont{M\'elin}(2006)}]{melin2006}
\bibinfo{author}{\bibfnamefont{R.}~\bibnamefont{M\'elin}},
  \bibinfo{journal}{Phys. Rev. B} \textbf{\bibinfo{volume}{73}},
  \bibinfo{pages}{174512} (\bibinfo{year}{2006}).

\bibitem[{\citenamefont{Duhot and M\'elin}(2006)}]{duhot2006}
\bibinfo{author}{\bibfnamefont{S.}~\bibnamefont{Duhot}} \bibnamefont{and}
  \bibinfo{author}{\bibfnamefont{R.}~\bibnamefont{M\'elin}},
  \bibinfo{journal}{Eur. Phys. J. B} \textbf{\bibinfo{volume}{53}},
  \bibinfo{pages}{257} (\bibinfo{year}{2006}).

\bibitem[{\citenamefont{Brinkman and Golubov}(2006)}]{brinkman2006}
\bibinfo{author}{\bibfnamefont{A.}~\bibnamefont{Brinkman}} \bibnamefont{and}
  \bibinfo{author}{\bibfnamefont{A.~A.} \bibnamefont{Golubov}},
  \bibinfo{journal}{Phys. Rev. B} \textbf{\bibinfo{volume}{74}},
  \bibinfo{pages}{214512} (\bibinfo{year}{2006}).

\bibitem[{\citenamefont{Levy~Yeyati et~al.}(2007)\citenamefont{Levy~Yeyati,
  Bergeret, Martin-Rodero, and Klapwijk}}]{yeyati2007}
\bibinfo{author}{\bibfnamefont{A.}~\bibnamefont{Levy~Yeyati}},
  \bibinfo{author}{\bibfnamefont{F.~S.} \bibnamefont{Bergeret}},
  \bibinfo{author}{\bibfnamefont{A.}~\bibnamefont{Martin-Rodero}},
  \bibnamefont{and} \bibinfo{author}{\bibfnamefont{T.~M.}
  \bibnamefont{Klapwijk}}, \bibinfo{journal}{Nature Physics}
  \textbf{\bibinfo{volume}{3}}, \bibinfo{pages}{455} (\bibinfo{year}{2007}).

\bibitem[{\citenamefont{Bardeen et~al.}(1957)\citenamefont{Bardeen, Cooper, and
  Schrieffer}}]{bardeen1957}
\bibinfo{author}{\bibfnamefont{J.}~\bibnamefont{Bardeen}},
  \bibinfo{author}{\bibfnamefont{L.~N.} \bibnamefont{Cooper}},
  \bibnamefont{and} \bibinfo{author}{\bibfnamefont{J.~R.}
  \bibnamefont{Schrieffer}}, \bibinfo{journal}{Phys. Rev.}
  \textbf{\bibinfo{volume}{108}}, \bibinfo{pages}{1175} (\bibinfo{year}{1957}).

\bibitem[{\citenamefont{Giaever}(1960)}]{giaever1960}
\bibinfo{author}{\bibfnamefont{I.}~\bibnamefont{Giaever}},
  \bibinfo{journal}{Phys. Rev. Lett.} \textbf{\bibinfo{volume}{5}},
  \bibinfo{pages}{147} (\bibinfo{year}{1960}).

\bibitem[{\citenamefont{Devoret et~al.}(1990)\citenamefont{Devoret, Esteve,
  Grabert, Ingold, Pothier, and Urbina}}]{devoret1990}
\bibinfo{author}{\bibfnamefont{M.~H.} \bibnamefont{Devoret}},
  \bibinfo{author}{\bibfnamefont{D.}~\bibnamefont{Esteve}},
  \bibinfo{author}{\bibfnamefont{H.}~\bibnamefont{Grabert}},
  \bibinfo{author}{\bibfnamefont{G.-L.} \bibnamefont{Ingold}},
  \bibinfo{author}{\bibfnamefont{H.}~\bibnamefont{Pothier}}, \bibnamefont{and}
  \bibinfo{author}{\bibfnamefont{C.}~\bibnamefont{Urbina}},
  \bibinfo{journal}{Phys. Rev. Lett.} \textbf{\bibinfo{volume}{64}},
  \bibinfo{pages}{1824} (\bibinfo{year}{1990}).

\bibitem[{\citenamefont{Kastalsky et~al.}(1991)\citenamefont{Kastalsky,
  Kleinsasser, Greene, Bhat, Milliken, and Harbison}}]{kastalsky1991}
\bibinfo{author}{\bibfnamefont{A.}~\bibnamefont{Kastalsky}},
  \bibinfo{author}{\bibfnamefont{A.~W.} \bibnamefont{Kleinsasser}},
  \bibinfo{author}{\bibfnamefont{L.~H.} \bibnamefont{Greene}},
  \bibinfo{author}{\bibfnamefont{R.}~\bibnamefont{Bhat}},
  \bibinfo{author}{\bibfnamefont{F.~P.} \bibnamefont{Milliken}},
  \bibnamefont{and} \bibinfo{author}{\bibfnamefont{J.~P.}
  \bibnamefont{Harbison}}, \bibinfo{journal}{Phys. Rev. Lett.}
  \textbf{\bibinfo{volume}{67}}, \bibinfo{pages}{3026} (\bibinfo{year}{1991}).

\bibitem[{\citenamefont{van Wees et~al.}(1992)\citenamefont{van Wees, de~Vries,
  Magn\'ee, and Klapwijk}}]{vanwees1992}
\bibinfo{author}{\bibfnamefont{B.~J.} \bibnamefont{van Wees}},
  \bibinfo{author}{\bibfnamefont{P.}~\bibnamefont{de~Vries}},
  \bibinfo{author}{\bibfnamefont{P.}~\bibnamefont{Magn\'ee}}, \bibnamefont{and}
  \bibinfo{author}{\bibfnamefont{T.~M.} \bibnamefont{Klapwijk}},
  \bibinfo{journal}{Phys. Rev. Lett.} \textbf{\bibinfo{volume}{69}},
  \bibinfo{pages}{510} (\bibinfo{year}{1992}).

\bibitem[{\citenamefont{Beenakker}(1992)}]{beenakker1992}
\bibinfo{author}{\bibfnamefont{C.~W.~J.} \bibnamefont{Beenakker}},
  \bibinfo{journal}{Phys. Rev. B} \textbf{\bibinfo{volume}{46}},
  \bibinfo{pages}{12841} (\bibinfo{year}{1992}).

\bibitem[{\citenamefont{Volkov and Klapwijk}(1992)}]{volkov1992}
\bibinfo{author}{\bibfnamefont{A.}~\bibnamefont{Volkov}} \bibnamefont{and}
  \bibinfo{author}{\bibfnamefont{T.}~\bibnamefont{Klapwijk}},
  \bibinfo{journal}{Phys. Lett. A} \textbf{\bibinfo{volume}{168}},
  \bibinfo{pages}{217} (\bibinfo{year}{1992}).

\bibitem[{\citenamefont{Volkov}(1993)}]{volkov1993}
\bibinfo{author}{\bibfnamefont{A.~F.} \bibnamefont{Volkov}},
  \bibinfo{journal}{Phys. Lett. A} \textbf{\bibinfo{volume}{174}},
  \bibinfo{pages}{144} (\bibinfo{year}{1993}).

\bibitem[{\citenamefont{Volkov et~al.}(1993)\citenamefont{Volkov, Zaitsev, and
  Klapwijk}}]{volkov1993b}
\bibinfo{author}{\bibfnamefont{A.}~\bibnamefont{Volkov}},
  \bibinfo{author}{\bibfnamefont{A.}~\bibnamefont{Zaitsev}}, \bibnamefont{and}
  \bibinfo{author}{\bibfnamefont{T.}~\bibnamefont{Klapwijk}},
  \bibinfo{journal}{Physica C} \textbf{\bibinfo{volume}{210}},
  \bibinfo{pages}{21} (\bibinfo{year}{1993}).

\bibitem[{\citenamefont{Volkov}(1994)}]{volkov1994}
\bibinfo{author}{\bibfnamefont{A.~F.} \bibnamefont{Volkov}},
  \bibinfo{journal}{Physica B} \textbf{\bibinfo{volume}{203}},
  \bibinfo{pages}{267} (\bibinfo{year}{1994}).

\bibitem[{\citenamefont{Schechter et~al.}(2001)\citenamefont{Schechter, Imry,
  and Levinson}}]{schechter2001}
\bibinfo{author}{\bibfnamefont{M.}~\bibnamefont{Schechter}},
  \bibinfo{author}{\bibfnamefont{Y.}~\bibnamefont{Imry}}, \bibnamefont{and}
  \bibinfo{author}{\bibfnamefont{Y.}~\bibnamefont{Levinson}},
  \bibinfo{journal}{Phys. Rev. B} \textbf{\bibinfo{volume}{64}},
  \bibinfo{pages}{224513} (\bibinfo{year}{2001}).

\bibitem[{\citenamefont{Ingold et~al.}(1994)\citenamefont{Ingold, Grabert, and
  Eberhardt}}]{ingold1994}
\bibinfo{author}{\bibfnamefont{G.-L.} \bibnamefont{Ingold}},
  \bibinfo{author}{\bibfnamefont{H.}~\bibnamefont{Grabert}}, \bibnamefont{and}
  \bibinfo{author}{\bibfnamefont{U.}~\bibnamefont{Eberhardt}},
  \bibinfo{journal}{Phys. Rev. B} \textbf{\bibinfo{volume}{50}},
  \bibinfo{pages}{395} (\bibinfo{year}{1994}).

\bibitem[{\citenamefont{Octavio et~al.}(1983)\citenamefont{Octavio, Tinkham,
  Blonder, and Klapwijk}}]{octavio1983}
\bibinfo{author}{\bibfnamefont{M.}~\bibnamefont{Octavio}},
  \bibinfo{author}{\bibfnamefont{M.}~\bibnamefont{Tinkham}},
  \bibinfo{author}{\bibfnamefont{G.~E.} \bibnamefont{Blonder}},
  \bibnamefont{and} \bibinfo{author}{\bibfnamefont{T.~M.}
  \bibnamefont{Klapwijk}}, \bibinfo{journal}{Phys. Rev. B}
  \textbf{\bibinfo{volume}{27}}, \bibinfo{pages}{6739} (\bibinfo{year}{1983}).

\bibitem[{\citenamefont{Feinberg}(2003)}]{feinberg2003}
\bibinfo{author}{\bibfnamefont{D.}~\bibnamefont{Feinberg}},
  \bibinfo{journal}{Eur. Phys. J. B} \textbf{\bibinfo{volume}{36}},
  \bibinfo{pages}{419} (\bibinfo{year}{2003}).

\bibitem[{\citenamefont{Golubev et~al.}(2009)\citenamefont{Golubev, Kalenkov,
  and Zaikin}}]{golubev2009}
\bibinfo{author}{\bibfnamefont{D.}~\bibnamefont{Golubev}},
  \bibinfo{author}{\bibfnamefont{M.}~\bibnamefont{Kalenkov}}, \bibnamefont{and}
  \bibinfo{author}{\bibfnamefont{A.}~\bibnamefont{Zaikin}},
  \bibinfo{journal}{Phys. Rev. Lett.} \textbf{\bibinfo{volume}{103}},
  \bibinfo{pages}{067006} (\bibinfo{year}{2009}).

\end{thebibliography}

\end{document}